\documentclass[12pt,preprint]{aastex}
\usepackage{natbib}
\usepackage{multirow}
\usepackage{amsmath}
\usepackage{graphicx}

\shorttitle{A \textit{Chandra} HETGS Observation of IRAS\,18325-5926}
\shortauthors{Mocz et al.}
\begin{document}
\title{A Detection of an X-ray Wind and an Ionized Disk in the \textit{Chandra} HETGS
Observation of the Seyfert 2 Galaxy IRAS\,18325-5926}

\author{Philip Mocz}
\affil{Harvard University, Cambridge, MA 02138}
\email{pmocz@fas.harvard.edu}
\and
\author{Julia C. Lee}
\affil{Harvard University, Department of Astronomy, 60 Garden
Street, Cambridge, MA 02138}
\affil{Harvard-Smithsonian Center for Astrophysics, 60 Garden
Street, Cambridge, MA 02138}
\email{jclee@cfa.harvard.edu}
\and
\author{Kazushi Iwasawa}
\affil{ICREA Research  
Professor at Institut de Ci\`encies del Cosmos, Universitat de  
Barcelona, Mart\'i i Franqu\`es, 1, 08028 Barcelona, Spain}

\email{kazushi.iwasawa@icc.ub.edu}
\and
\author{Claude R. Canizares}
\affil{Kavli Institute for Astrophysics and Space Research, Massachusetts Institute
of Technology, Cambridge, MA 02139}
\email{crc@mit.edu}

ApJ accepted 31 December 2010.


\begin{abstract}
We analyze the \textit{Chandra} High Energy Transmission Grating Spectrometer (HETGS) observation of the Seyfert $2$ AGN IRAS\,18325-5926. We detect a $v=-360^{+41}_{-66}$~km~s$^{-1}$ blueshifted ionized absorber in the X-ray spectrum, with photoionization parameter $\log\xi=2.0^{+0.1}_{-0.1}$ and hydrogen column density $N_{\rm H}=1.55^{+0.75}_{-0.38}\times10^{21}$~cm$^{-2}$. The absorber may be a photoionized wind originating in the obscuring torus/global covering around the black hole or outer edge of the accretion disk. The estimated mass outflow rate suggests that the supermassive black hole in IRAS\,18325-5926 may significantly affect the large-scale environment of the host galaxy, unless the solid angle subtended by the outflow or the gas filling factor is small. A second warm absorber may be needed to explain the absorption features in the vicinity of the iron K edge, although insufficient counts in the data beyond $7.0$~keV make it difficult to assess the nature of the second absorber. Most plausible is a high ionization ($\log\xi\sim 2.3$ to $2.6$), high column density ($N_{\rm H}\sim10^{23}$~cm$^{-2}$) absorber with $v\sim -3000$ to $0$~km~s$^{-1}$, although these parameters are not well constrained. We also examine the broad Fe K emission line in the spectrum, which is likely due to Fe XXV in a highly ionized accretion disk inclined at $25^\circ$, consistent with the \textit{XMM-Newton} EPIC observations of the emission line. Because we are able to view both the obscuring gas and the accretion disk of IRAS\,18325-5926, the surrounding gas of IRAS\,18325-5926 may be patchy or we are viewing the system at an angle just grazing the obscuring torus.
\end{abstract}
\keywords{galaxies: active -- galaxies: individual (IRAS 18325-5925) -- galaxies: winds -- X-rays: galaxies}

\section{Introduction}\label{Introduction}
IRAS\,18325-5926 is a Seyfert $2$ type galaxy with a redshift of $z=0.01982\pm0.0006$ \citep{1995AJ....110..551I, 1996MNRAS.279..837I, 2004MNRAS.347..411I}. The redshift was determined from the narrow, low ionization line ([OII]), which is not affected by a blueshifted component which is seen in [OIII] and the Balmer lines in the optical band \citep{1995AJ....110..551I}.
Resolved spectral features from previous studies include a broad Fe K$\alpha$ emission line at $\sim 6.7$~keV, thought to arise because of 
Compton scattering and relativistic blurring of a highly ionized disk, and a S XVI radiative recombination continuum (RRC) feature at $\sim 3.4$~keV, which supports the idea of a reflection model from an ionized disk.
This study by \citet{2004MNRAS.347..411I} was carried out using a compendium of X-ray satellites (\textit{Ginga}, \textit{ASCA}, \textit{RXTE}, \textit{BeppoSAX}, \textit{XMM-Newton}). 
Since the last publication on IRAS\,18325-5926 was by \citet{2004MNRAS.347..411I}, hereafter I04, this paper will draw comparisons primarily with that paper.

IRAS\,18325-5926 is unique because despite being classified as a Seyfert $2$ type galaxy, it shares some characteristics with Seyfert~$1$s in the hard ($>2$~keV) X-ray spectrum. The narrow emission-line dominated optical spectrum and the steep continuum slope $\Gamma=2.2$ suggest IRAS\,18325-5926 is a Seyfert $2$ type galaxy \citep{1995AJ....110..551I, 1998MNRAS.295L..20I}. But the absorption-free hard X-ray spectrum, high variability in the X-ray source, and the moderately absorbed column density of $10^{22}$~cm$^{-2}$ are characteristics more similar to an obscured Seyfert~$1$ type galaxy \citep{1998MNRAS.295L..20I}.
IRAS\,18325-5926 may not have an optically thick torus, as normally envisaged for Seyfert galaxies and the Unified Model \citep{1977ApJ...213..635R,1987PASP...99..309L,1993ARA&A..31..473A}, but rather obscuration in terms of a global covering may be preferred, since the degree of extinction in the X-ray and optical are comparable \citep{1995AJ....110..551I}.
IRAS\,18325-5926 also lacks a narrow $6.4$~keV emission line that accompanies the reflection spectrum from a torus usually seen in Seyfert galaxies (see also \citet{1996MNRAS.282.1038I} for a discussion of MCG-6-30-15, somewhat similar to IRAS\,18325-5926).

The goal of the present work is to identify finer spectral details in the spectrum of IRAS\,18325-5926 in data gathered by the \textit{Chandra} High Energy Transmission Grating Spectrometer \citep[HETGS;][]{2005PASP..117.1144C} in an effort to gain a deeper understanding of the ionization and kinematics of the surrounding X-ray absorbing material. High spectral resolution observations with \textit{Chandra} enable the detection of narrower absorption and emission lines superimposed on the continuum of the source in order that we may better probe dynamics and plasma conditions. Absorption features may indicate the presence of substantial amounts of ionized gas which are bathed in radiation by the central source; observed blueshifts will further tell us that material is moving away from the AGN in an X-ray wind. This high resolution study,
coupled with additional studies (with \textit{RXTE}, \textit{BeppoSAX}, \textit{XMM-Newton}) of the broadened 
emission lines can help us to better understand the geometry of IRAS\,18325-5926, as regards our line-of-sight viewing, its accretion disk properties as well as surrounding environment, and influences. 

\section{Observations and Analysis}\label{ObservationsandAnalysis}
\begin{figure}[htbp]
\centering
\includegraphics[width=1.0\textwidth]{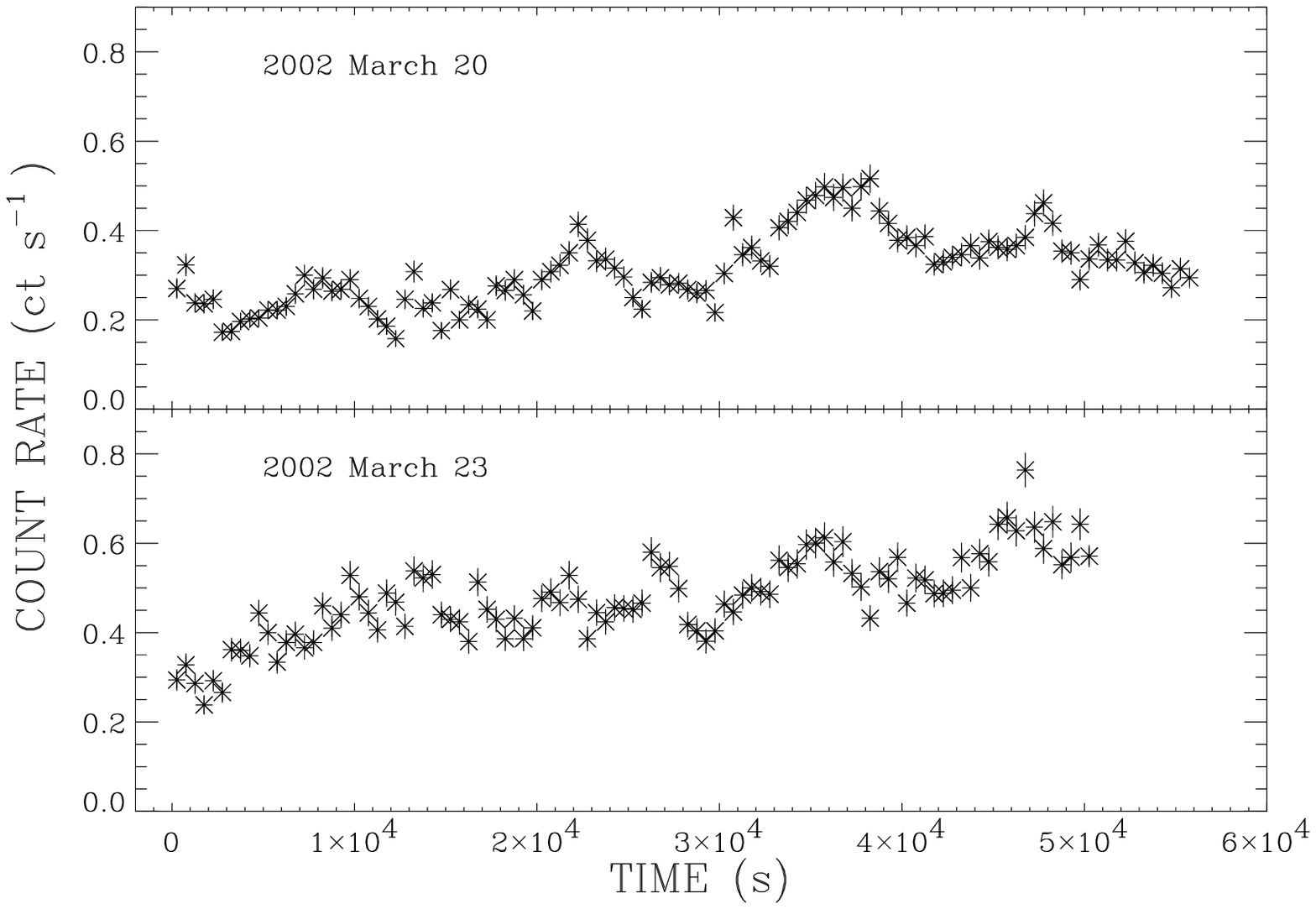}
\caption{The IRAS\,18325-5926 \textit{Chandra} ACIS-S HETGS light curve (excluding the $0$th order) binned at $500$-s intervals.}
\label{cntsT}
\end{figure}

IRAS\,18325-5926 was observed on $2002$ March $20$ (Obs ID $3148$) and on $2002$ March $23$ (Obs ID $3452$) 
with the ACIS-S (Advanced CCD Imaging Spectrometer - Spectral component) instrument coupled with the HETGS on \textit{Chandra}. 
The exposure times of the two observations were $56.9$~ks and $51.09$~ks respectively. The light-curve count rate of the dispersed data fluctuated between a low of $\sim0.1$~cts~s$^{-1}$ to a high of $\sim0.8$~cts~s$^{-1}$ (Fig.~\ref{cntsT}). The data were reprocessed with CIAO~v~$4.1.2$ and new level $2$ event files were created. The $+1$ and $-1$ orders of the HEG and MEG grating arms were combined, as were both the March $20$ and $23$ observations. For Fe emission line analysis, we also investigated the \textit{XMM-Newton} observations taken a year earlier.

Data was binned by a factor of $4$ (HETGS resolution). There are enough counts to decisively study absorption lines in the $1.0$--$7.0$~keV spectral band. This region contains counts between $5$ and $40$ per bin. 
The signal-to-noise ratio is at least $3:1$ in this region, but is much lower outside this range. We analyze the MEG spectrum from $0.8$--$1.2$~keV, and the HEG spectrum from $1.2$--$8.0$~keV (extending slightly into the regions of lower counts), and we cross-check absorption features in the HEG and MEG spectra between $1.2$--$2.0$~keV. Low counts ($<5$ counts per bin) are seen in the MEG data from $0.8$--$1.0$~keV and in the HEG data from $1.2$--$1.3$~keV and $7.0$--$8.0$~keV.

We use the Interactive Spectral Interpretation System (ISIS) \citep{2002hrxs.confE..17H} to fit the spectra. Absorption features are found by fitting the spectrum with an ionized (warm) absorber, {\sc warmabs}, using the photoionization code {\sc xstar}\footnote{available at http://starchild.gsfc.nasa.gov/lheasoft/xstar/xstar.html} \citep{2001ApJS..134..139B}.  
The ionization parameter $\xi$, column density, and outflow velocity are variable parameters in the model which are optimized for best-fit. All spectral fits account for the line-of-sight Galactic column $N_{\rm H}=7.4\times 10^{20} $~cm$^{-2}$. The {\sc tbabs} model \citep{2006HEAD....9.1360W} which corrects for X-ray absorption due to gas-phase and grain-phase ISM, and molecules in the ISM, is applied for this purpose.

A gap in the detector chip, where the effective area is small, was noticed in the $+1$ order of the HEG grating arm in both observations of IRAS\,18325-5926. The $-1$ grating arm appeared normal. The spectral band affected is between $2.5$--$2.7$~keV. This region is not reliable for the detection of absorption features in our source, and therefore is ignored for fitting.

\subsection{Broadband Continuum}\label{BroadbandContinuum}

\begin{figure}[htbp]
\centering
\includegraphics[height=1.0\textwidth,angle=270.0]{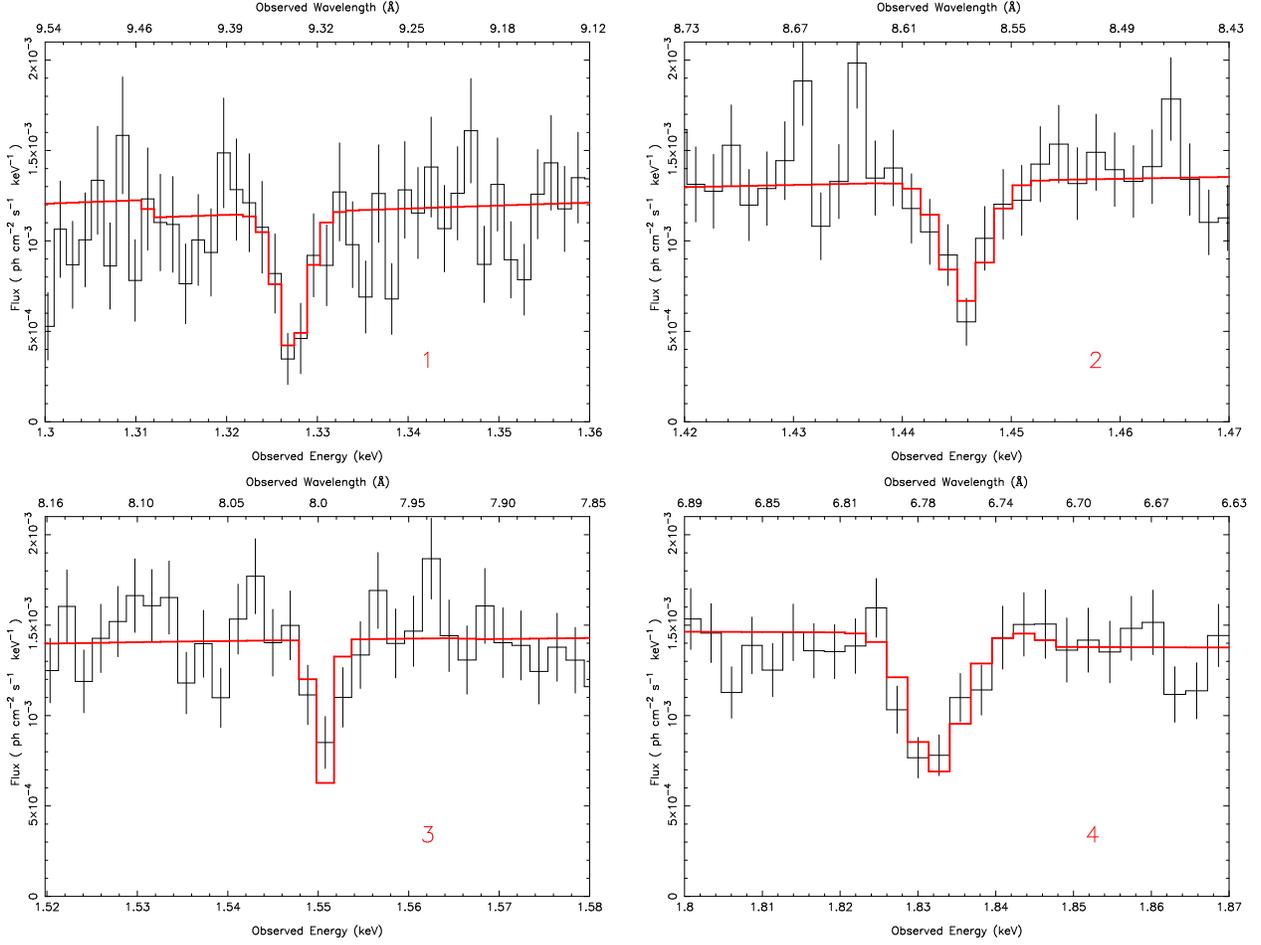}
\caption{Strong absorption features seen in the soft X-ray HEG spectrum identified by the $-360$~km~s$^{-1}$ ionized absorber. 
[1] (ID\#~$9$, Mg {\sc XI})
[2] (ID\#~$11$, Mg {\sc XII})
[3] (ID\#~$10$, Mg {\sc XI})
[4] (ID\#~$12$, Si {\sc XIII})}
\label{abs}
\end{figure}

The X-ray continuum measured at the epoch of our \textit{Chandra} observation is best modeled by a power-law photon-index $\Gamma=2.01^{+0.07}_{-0.10}$ modified by a partial absorber of $N_{\rm H}=1.33^{+0.02}_{-0.03}\times10^{22}$~cm$^{-2}$ and $0.94^{+0.01}_{-0.01}$ covering fraction, and a line-of-sight Galactic column of $N_{\rm H}=7.4\times 10^{20} $~cm$^{-2}$. $90\%$ confidence errors are reported. 
In addition we also find that a broad Fe XXV emission line (\S~\ref{FeKEmissionLine}) and an ionized (``warm'') absorber intrinsic to the source (\S~\ref{NarrowAbsorptionLines}), are needed to describe some of the spectral details.
The best-fit model parameters are summarized in Table~\ref{par}.
Strong absorption lines are clearly detected (see Fig.~\ref{abs}).

Historically, the best-fit photon index for IRAS\,18325-5926, fit along with a partial absorber, has ranged from $\Gamma=2.00\pm 0.05$ to $2.26\pm0.05$ \citep[][I04]{1995AJ....110..551I, 1996MNRAS.279..837I}. We find, fitting with a similar continuum, that the high-resolution \textit{Chandra} spectrum has a best-fit $\Gamma=2.01^{+0.07}_{-0.10}$, at the lower end of previously published values, although fits to \textit{Chandra} alone have typically given a slightly lower $\Gamma$ for a number of sources. Also, many of the previous observations have fit to X-ray spectra above $2$~keV, whereas we include the spectrum between $0.8$--$1.0$~keV for fitting. If we try to fit the continuum of the \textit{Chandra} observation for energies greater than $2$~keV, then we find a best-fit photon index of $\Gamma\sim2.1$, consistent with previous measurements.
Partial $95\%$ covering fraction absorption and $N_{\rm H}=1.3^{+0.1}_{-0.2}\times10^{22}$~cm$^{-2}$ column density are also required and agree with previous I04 values. 
The partial covering absorber (with covering fraction $f=0.95$) makes most sense in the context of a scattering model: $5\%$ of the continuum emission is electron-scattered around the absorber to produce the observed spectrum.

The ($2$--$10$~keV) unabsorbed X-ray luminosity calculated for IRAS\,18325-5926 at the epoch of our observation is $L_{\rm x}=1.79\times 10^{43}$~erg s$^{-1}$ (based on a Hubble constant of $H_0=72$~km~s$^{-1}$ Mpc$^{-1}$; \citealt{2001ApJ...553...47F}). The $1$--$1000$ Rydberg ($13.6$~eV-$13.6$~keV) luminosity is estimated as $L=2.08\times 10^{43}$~erg~s$^{-1}$.
The $2$--$10$~keV flux of our observation is $F_{2-10}=2.23\times 10^{-11}$~erg~cm$^{-2}$~s$^{-1}$, best corresponding to the $2000$ March $31$ \textit{BeppoSAX}
$F_{\rm{SAX,}2\rm{-}10}=2.0\times 10^{-11}$~erg~cm$^{-2}$~s$^{-1}$ observation of IRAS\,18325-5926 reported by I04; else the $2$--$10$~keV fluxes of IRAS\,18325-5926 have historically varied between $(1.2$--$2.6)\times 10^{-11}$~erg~cm$^{-2}$~s$^{-1}$. 

The S XVI radiative recombination continuum (RRC) feature at $\sim 3.4$~keV detected in the I04 analysis of the \textit{RXTE} and \textit{XMM-Newton} X-ray spectra is not detected in the \textit{Chandra} spectrum. 

\subsection{Fe K$\alpha$ Emission Line}\label{FeKEmissionLine}

A broad line has been reported on numerous occasions for IRAS\,18325-5926 (e.g., \textit{ASCA}: \citet{1996MNRAS.279..837I}, I04,
\textit{Ginga}: I04,
\textit{RXTE}: I04,
\textit{BeppoSAX}: I04). 
For a number of AGN, the Fe K$\alpha$ emission line is thought to arise from the reprocessing of X-ray radiation by iron fluorescence from a relativistic accretion disk \citep[][and references therein]{2003PhR...377..389R}. 
The line broadening due to strong gravitational effects near the center of the source may have a double-peaked profile, or the double peak may be smeared out. The line emission from a highly ionized disk can also broaden to some degree by Compton scattering. Both mechanisms may be needed to fully explain the observed features, since, as noted by I04 the broad line in IRAS\,18325-5926 is thought to be mainly due to Fe XXV (lab frame emission at $6.7$~keV) in a highly ionized accretion disk that shows reflection and relativistic blurring.
As also noted by I04, Compton scattering broadens the emission line to some degree, but it cannot 
explain all of the broadening, so that additional relativistic broadening is required. 

In order to better investigate the broad Fe emission line to compare with previous results, we consider it with heavily binned ($\times 16$, $0.04$~\AA) \textit{Chandra} data to approximately match the resolution of previous lower resolution instruments, e.g. \textit{Ginga}, \textit{ASCA}, \textit{RXTE}, and \textit{BeppoSAX}.
We model the Fe K$\alpha$ line with the {\sc diskline} model (see \citealt{1989MNRAS.238..729F}) to account for the skewed shape towards lower energies as a consequence of relativistic blurring of the emission line. The {\sc diskline} model describes a relativistically blurred emission line around a Schwarzschild (stationary) black hole. The inner and outer radii are fixed at $6$ and $200$ gravitational radii respectively, as in I04, while the emissivity ($\beta^2$), normalization, and inclination are left as free parameters.
The disk is found to be inclined at approximately ${24.4^\circ}^{+6.6}_{-10.5}$, consistent with the results of I04,
based on fits to \textit{Ginga}, \textit{ASCA}, \textit{RXTE}, \textit{BeppoSAX}, and \textit{XMM-Newton}, where the emission line was modeled with an inclination of $25.84^\circ$. In an older paper that investigates \textit{ASCA} data and considers the emission features to be due to a cold accretion disk, the inclination of the disk is found to be higher, at $40^\circ$--$50^\circ$ \citep{1996MNRAS.279..837I}, 
since in the cold accretion disk model, neutral fluorescent Fe K$\alpha$ emission at $6.4$~keV needs a large Doppler blueshift from a higher inclination disk to move the blue horn of the cold line to $6.7$~keV.
Here, with better \textit{Chandra} HETGS spectra, we measure the emission line to originate from an ionized disk giving rise to $6.7$~keV emission from He-like Fe.

A simple Gaussian fit can also be applied to the broad emission line for the heavily binned \textit{Chandra} spectrum, although the line profile is visibly skewed. The best-fit Gaussian is centered at the observed energy $6.49^{+0.22}_{-0.38}$~keV (de-redshifted energy $6.62$~keV), which is higher than the $6.4$~keV (lab-frame) emission of neutral fluorescent Fe K$\alpha$. The corresponding line flux and velocity width are $F_{{\rm Fe}25}=5.7^{+3.3}_{-2.0}\times 10^{-5}$~erg~cm$^{-2}$~s$^{-1}$ and $\sigma=0.47^{+0.42}_{-0.14}$~keV respectively, corresponding to a velocity broadening of $45400$~km~s$^{-1}$. The line has an equivalent width (EW) of $254$~eV.

\begin{figure}[htbp]
\centering
\includegraphics[width=1.0\textwidth]{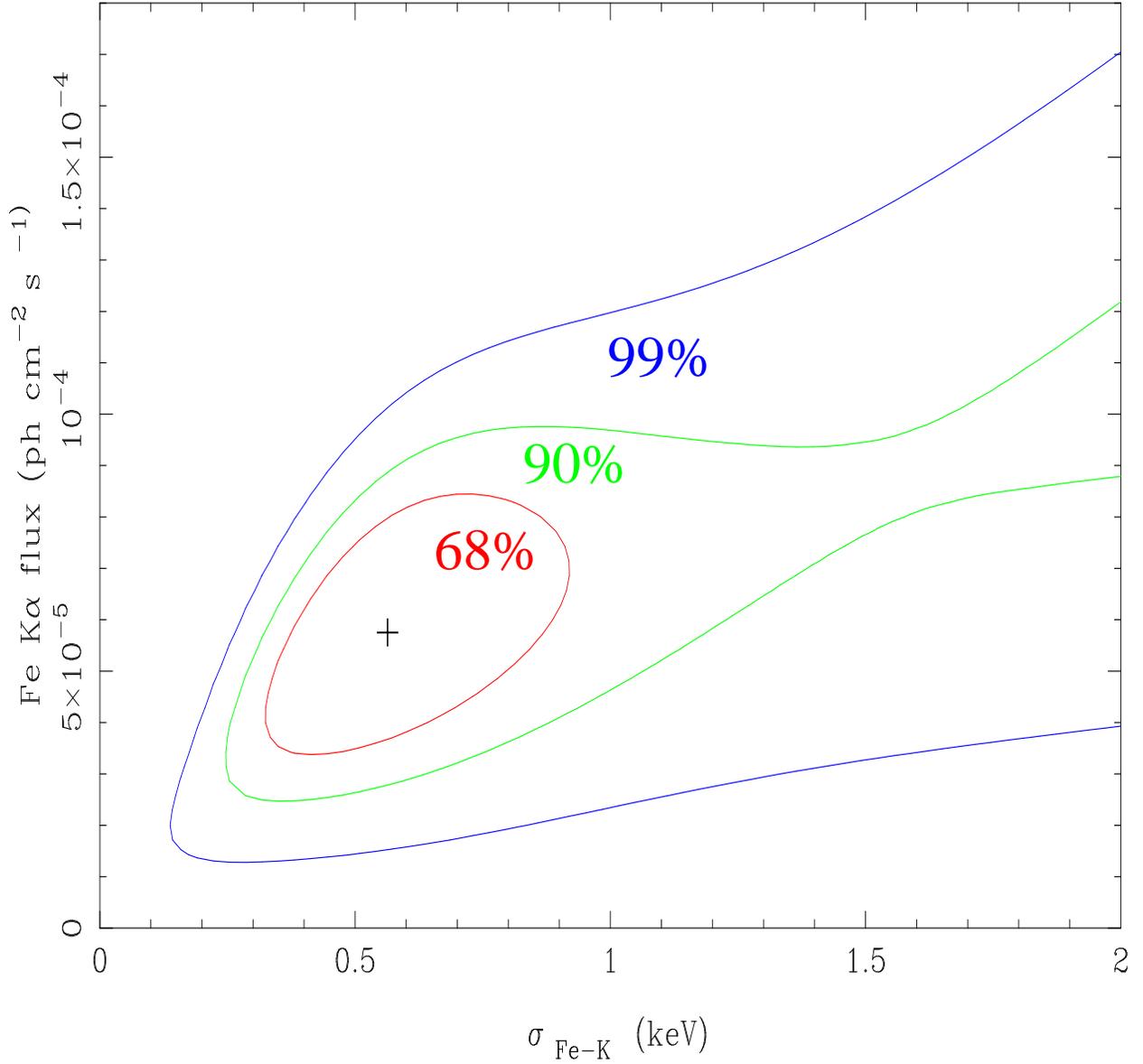}
\caption{The $60\%$, $90\%$, and $99\%$ confidence limits on the narrow component of the broad iron line seen in IRAS\,18325-5926 shows that a broad line is preferred by the data.}
\label{contour}
\end{figure}

\textit{Chandra} is ideal for measuring the narrow component of the broad emission line due to the high spectral resolution of the HETGS. 
We fit the emission line using spectral binning consistent with the HEG capabilities (binned $\times 4$, $0.01$~\AA) and find that the rest energy of the emission line is likely at $E=6.7$~keV, consistent with emission from Fe XXV. We arrive at this value by fitting the emission line with a broad and narrow Gaussian (no {\sc diskline}).
The central peak of the broad component (spectra binned to \textit{Chandra} resolution) is at the observed energy $6.54^{+0.16}_{-0.39}$~keV (de-redshifted energy $6.67$~keV) and has a velocity width of  $\sigma=0.45^{+0.17}_{-0.15}$~keV (corresponding to a full width at half maximum (FWHM) of $43700$~km~s$^{-1}$). The flux is $F_{{\rm Fe}25}=5.2^{+3.8}_{-1.5}\times 10^{-5}$~erg~cm$^{-2}$~s$^{-1}$ and the equivalent width of the line is $236$~eV. 
These values agree with the values for the broad Gaussian fit over the heavily binned spectra.
In previous studies, the broad iron K line has been modeled with Gaussians found to have centers at $6.38$--$6.65$~keV (energies corrected for the redshift of the galaxy) and EWs were mostly between $200$--$300$~eV, although as high as $626$~eV in one observation (I04).
The Gaussian that best describes the narrow component in the \textit{Chandra} data has its peak at $6.59^{+0.20}_{-0.04}$~keV, corresponding to $6.72$~keV (He-like Fe) in the lab frame. The line has $\sigma=3.0^{+400}_{-3}$~eV (FWHM of $320$~km~s$^{-1}$), $F_{{\rm Fe}25}=3.9^{+6.1}_{-3.9}\times 10^{-6}$~erg~cm$^{-2}$~s$^{-1}$, and an EW of $30$~eV. 
It appears that the fit prefers just a broad line, with no narrow core. The narrow core in the data is not visibly strong.  Fig.~\ref{contour}, which shows $68$, $90$, and $99\%$ confidence contours of the possible width of the emission line also support this. 

We also repeat the fit of the {\sc diskline} model to the spectra binned at HEG capability and find a similar fit as for the heavily binned spectrum. The disk inclination is ${25.8^\circ}^{+3.5}_{-7.6}$.
We also fit the Fe K$\alpha$ emission with the {\sc laor} \citep{1991ApJ...376...90L} model for a relativistically blurred emission line around a Kerr (rotating) black hole, which produces a comparable fit (similar chi-square). The two models are compared in Table~\ref{par2}.

\begin{figure}[htbp]
\centering
\includegraphics[height=1.0\textwidth,angle=270.0]{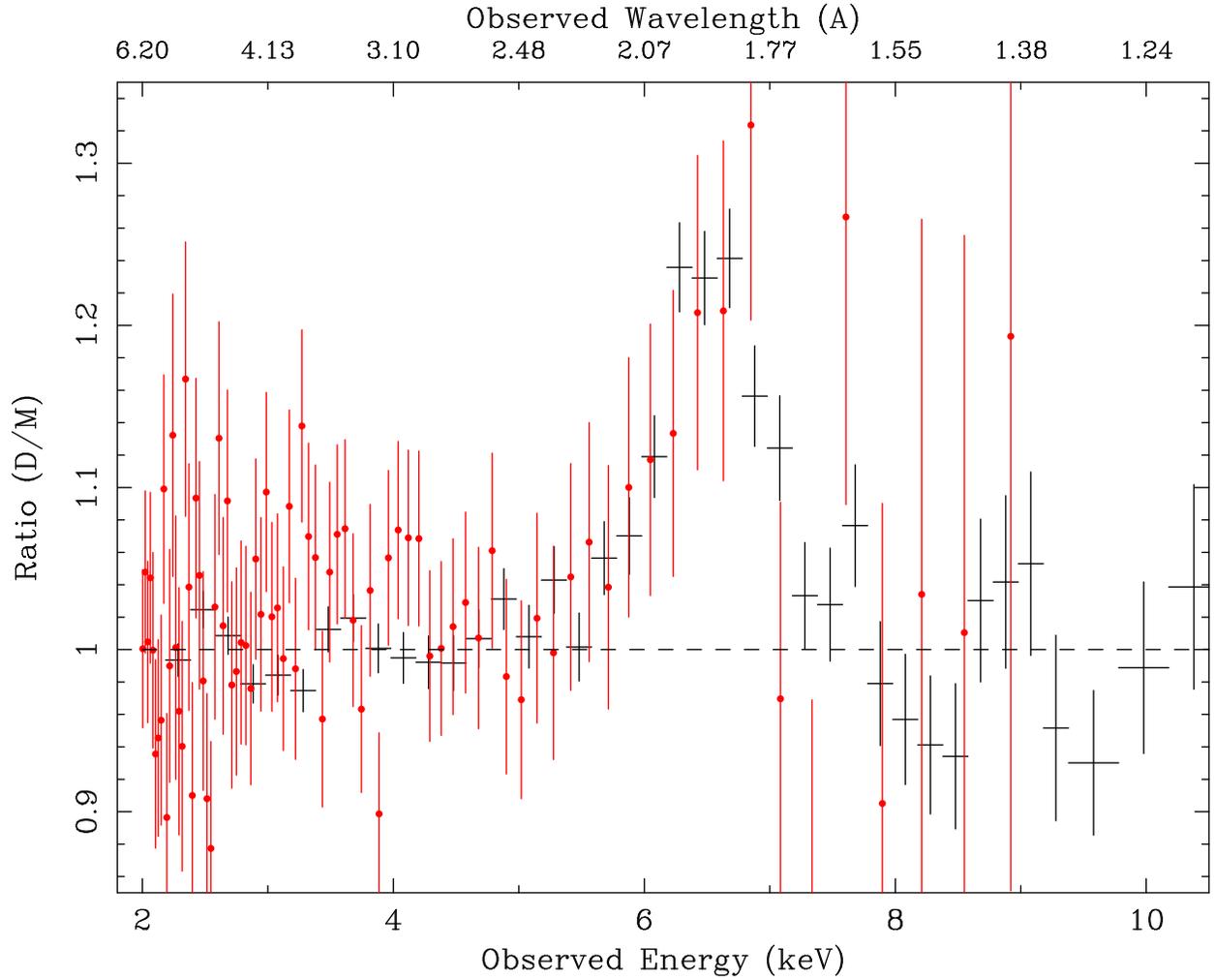}
\caption{The Iron K$\alpha$ emission line. Ratio plot of data divided by best-fit continuum model for \textit{XMM-Newton} (black) and \textit{Chandra} spectra (red) show very similar profiles for a source flux which differed by a factor of two between the two observations, supporting the disk origin interpretation of the emission line.}
\label{xmm}
\end{figure}

We also investigate potential line changes over observations at different epochs. The Fe K$\alpha$ emission line plotted as a ratio against the continuum is presented in Fig.~\ref{xmm}.
In the figure, the \textit{Chandra} observation of the emission line is compared with an \textit{XMM-Newton} observation taken $1$ year earlier. The equivalent width of the broad line is comparable between the epochs of the \textit{Chandra} and \textit{XMM-Newton} observations, even though the source flux differed by a factor of two (the \textit{XMM-Newton} EW is $0.242$~keV and the flux is $3.0\times 10^{-5}$~erg~cm$^{-2}$~s$^{-1}$; I04). 

We also attempt to model the spectrum of IRAS\,18325-5926 including the broad Fe K$\alpha$ line in an alternative way, with a relativistically-blurred reflection model, as in \citet{2008MNRAS.391.2003Z}. 
The model\footnote{phabs(1)*(zpowerlw(1)+kdblur(1,reflionx)) in ISIS, reflionx is an additive table model} combines a power-law from the illuminating continuum with an optically-thick, constant density, ionized reflection model, {\sc reflionx} of \citet{2005MNRAS.358..211R} blurred with {\sc kdblur} of \citet{1991ApJ...376...90L} to account for relativistic effects due to strong gravity in the vicinity of the black hole affecting both the line and continuum.
The best-fit parameters are presented in Table~\ref{par2}. The 
chi-square ($\chi^2=1247$ for $1130$ d.o.f.) is slightly higher than for the {\sc diskline} and {\sc laor} models, and the model predicts significantly more line-of-sight absorption than accounted for by the galactic column ($N_{\rm H}=7.4\times 10^{20} $~cm$^{-2}$). 
While it is hard to distinguish between the models in a chi-square sense, the reflionx fits has the advantage that it self-consistently treats the reflection spectrum (continuum and emission line) and therefore is more physical than our other models, where line and continuum are treated as two separate model components.

\subsection{Narrow Absorption Lines}\label{NarrowAbsorptionLines}

Narrow \textit{absorption} features are also detected in the high-resolution \textit{Chandra} spectrum of IRAS\,18325-5926. 
The strongest absorption features in the soft X-ray spectrum ($<2$~keV), which appear in both the HEG and MEG data, are located at approximately observed energies $1.328$~keV ($9.340$~\AA) (ID\#~$9$, Mg {\sc XI}), $1.445$~keV ($8.578$~\AA) (ID\#~$11$, Mg {\sc XII}), $1.550$~keV ($7.997$~\AA) (ID\#~$10$, Mg {\sc XI}), $1.831$~keV ($6.772$~\AA) (ID\#~$12$, Si {\sc XIII}), and $1.969$~keV ($6.297$~\AA) (ID\#~$13$, Si {\sc XIV}), corresponding to He- and H-like transitions of magnesium and silicon (ID numbers correspond to numbers assigned to absorption features in Table~\ref{lines}). See also Fig.~\ref{abs}.

Many of the strongest features including the aforementioned strong lines are best-described by an ionized absorber blueshifted at $-360^{+41}_{-66}$~km~s$^{-1}$, relative to the systemic velocity of the source. The absorber, hereafter WA1, has a best-fit photoionization parameter of $\log\xi=2.01^{+0.07}_{-0.1}$ and column density $N_{\rm H}=1.6\times 10^{21}$~cm$^{-2}$, assuming solar abundances, as determined from an {\sc XSTAR} \citep{2004ApJS..155..675K} $\Gamma\sim 2$ ionizing continuum. Table~\ref{lines} lists the detected strong lines based on the best fit absorber, as determined by photoionization modeling with {\sc XSTAR}.

We also observe a very strong absorption feature at $2.54$~keV ($\sim 4.88$\AA), which is likely an artifact due to a gap at $2.5$~keV-$2.7$~keV in the $+1$ order HEG spectra as mentioned in \S~\ref{ObservationsandAnalysis}. This dip is not noticeably strong in the $-1$ order of the spectra. There are no plausible candidates with reasonable velocity shifts for an absorption feature at this energy. 

Nevertheless, we attempted to model the $2.54$~keV line on the basis that it is real and found that 
H-like sulfur (S XVI, rest frame energy $2.6$~keV) redshifted at $+3000$ to $+5300$~km~s$^{-1}$ with respect to the rest frame of the source (ionization parameter $\log\xi\geq 2.9$) is able to explain the feature (a redshift at $\sim3000$~km~s$^{-1}$ will place the predicted line at the observed peak of the $2.54$~keV line, however the line is wide so that a range of redshifts may work). However, such an absorber 
will also produce strong silicon lines where none are seen, thus requiring a significantly decreased silicon abundance relative to the solar value to explain the spectrum. The absorber does, however, accurately predict strong blips observed at $6.4$--$7.0$~keV in the spectra (see also \S~\ref{VicinityoftheFeKEdge} where we discuss this scenario again). Redshifted absorbers are less likely to be detected than blueshifted absorbers, although redshifted iron emission lines have been detected in other sources such as NGC~3516 (e.g. \citealt{2004MNRAS.355.1073I}), which are attributed to a localized flare illuminating a receding spot on the accretion disk. A strong dip in the MEG spectrum at $2.54$~keV is only present in the $+1$ order of the second observation of IRAS\,18325-5926 (Obs ID $3452$).

\subsection{Vicinity of the Fe K Edge: Evidence for A Second Absorber?}\label{VicinityoftheFeKEdge}

We observe a broad trough between $7$--$8$~keV and strong absorption at $6.42$~keV (observed energies), which have not been noted in previous observations of IRAS\,18325-5926. 
The $-360$~km~s$^{-1}$ warm absorber (WA1) we used to describe the strongest narrow absorption features does not predict any strong absorption at $6.42$~keV or from $7$--$8$~keV, and our best fit diskline, laor, and reflionx models do not explain these features.
Thus we investigate this region with a possible second absorber component. 
Since the disk is ionized sufficiently to produce a strong Fe XXV emission line, strong Compton broadening of the line is unavoidable and therefore we use the full ionized/smeared reflection fit to interpret that structure around the iron-edge. 
Compared with a simple diskline type fit, the best-fit reflionx fit did require significantly more absorption.
The reason for this is that the self-consistent iron line profile (which is strongly Compton broadened in addition to being relativistically smeared) has a blue wing which extends above $7$~keV and hence the model is compensating (i.e. removing that extra flux) by increasing the absorption. Nevertheless the trough feature is not fully explained by the reflionx model, an additional warm absorber may be needed.

We thoroughly explore a parameter space for this possible second absorber with photoionization parameter $\log\xi$ ranging between $-1$ and $4$ and blueshift velocities from $0$ to $-7000$~km~s$^{-1}$. We also explore the possibility of a redshifted absorber ($0$~km~s$^{-1}$ to $+5000$~km~s$^{-1}$), as well as Galactic ISM absorption (in which case we use the {\sc ismabs} model). 

The trough between $7$--$8$~keV and strong absorption at $6.42$~keV (observed energies) could be due to the complex profiles of iron K absorption; in the rest frame, a series of Fe K lines can appear at $\sim 6.4$--$6.5$~keV and at $\sim7.0$--$8.0$~keV for photoionization $\log\xi\geq 1.5$ 
as noted by the theoretical calculations of \citet{2004ApJS..155..675K}.
It turns out that it is difficult to fit this portion of our data as signal-to-noise drops to less than $3:1$ beyond $7$~keV and counts may be less than $3$ photons per bin. The reduced chi-squared values of models fit to the data do not change significantly when the trough and lines at $6$--$7$~keV are ignored, if we are fitting the entire energy spectrum.
Attempting to fit just the $6$--$7$~keV with chi-squared or Cash-statistic (ideal for low number of counts) fitting leads to very wide confidence intervals.
 However, these features in the vicinity of the Fe~K edge are reasonably plausibly real as the \textit{XMM-Newton} data also show a clear trough; a joint analysis of \textit{Chandra} and \textit{XMM-Newton}, however, would not be useful because of variability. 
 In addition, as we have mentioned, the model that used reflionx to explain the iron line needed a high amount of absorption, predicting that the blue wing of the iron emission line in this region should be suffering from absorption in this region. 
We find that a wide range of parameters for an absorber can describe the broad trough and strong $6.42$~keV absorption and there is not enough statistical significance to pick a single best-fit second warm absorber component over another. 
Thus, we fix various photoionizations for our second absorber ($\log\xi$ between $-1$ and $4$), choose the redshift/blushift accordingly so that the absorption predicted by the model falls at $6.42$~keV and at $7$--$8$~keV, and asses (by eye) what column densities could work to explain the vicinity of the Fe K edge and whether the model agrees well with the rest of the spectrum.

\begin{figure}[htbp]
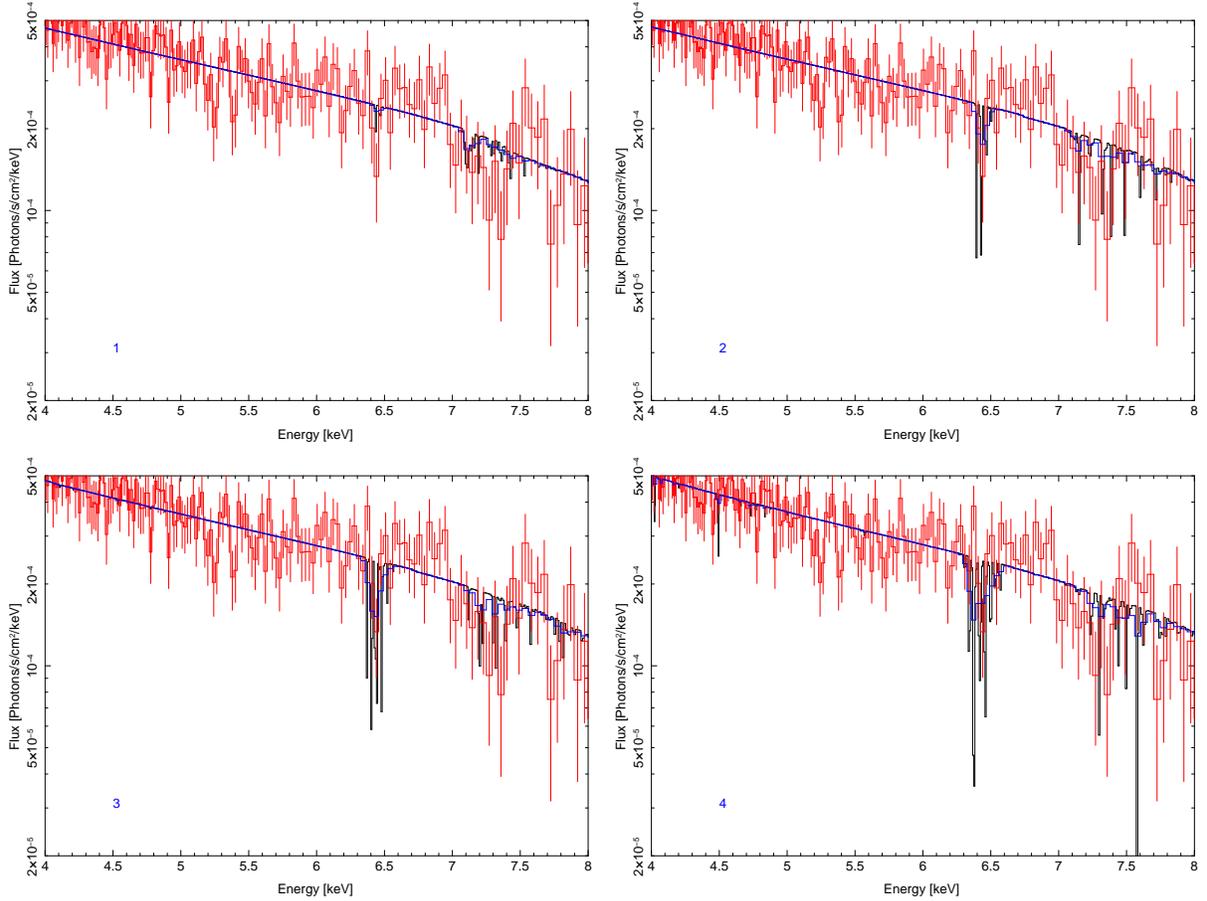

\centering
\includegraphics[height=0.48\textwidth,angle=270.0]{f7.eps}
\includegraphics[height=0.48\textwidth,angle=270.0]{f8.eps}
\includegraphics[height=0.48\textwidth,angle=270.0]{f9.eps}
\includegraphics[height=0.48\textwidth,angle=270.0]{f10.eps}
\caption{Exploring the parameter space to model the vicinity of the iron K edge. Due to the low number of counts in the data at $7$--$8$~keV, several warm absorbers produced reasonable fits to explain absorption features near the iron K edge. Due to noise in the spectrum, not enough absorption features were seen at lower energies to deduce a best-fit component. However, if we assume a non-redshifted warm absorber explains the features, then a photoionization, $\log\xi$, between $1.7$ and $2.6$ and velocity (rest frame of source) between $-60000$ and $0$~km~s$^{-1}$ are required.
[1]: $\log\xi=1.7$, $v=-6000$~km~s$^{-1}$.
[2]: $\log\xi=2.0$, $v=-4000$~km~s$^{-1}$.
[3]: $\log\xi=2.3$, $v=-3000$~km~s$^{-1}$.
[4]: $\log\xi=2.6$, $v=0$~km~s$^{-1}$.
$N_{\rm H}\sim 10^{23}$~cm$^{-2}$ in the shown fits. Models are shown in black, convolved models are overplotted in blue.}
\label{ke}
\end{figure}

 The first scenario we explore is the possibility that the features in the vicinity of the Fe K edge are due to a blueshifted warm absorber. We step through the photoionization parameter in $\Delta\log\xi=0.10$ steps and determine what velocity shift aligns predicted and observed features the best. We find that a photoionization parameter $\log\xi$ between $1.7$ and $2.6$  will predict absorptions at both $6.42$~keV and $7.0$--$8.0$~keV.
The exploration of parameter space is presented in Fig.~\ref{ke}. We summarize the findings below.

(1) If we set the  photoionization $\log\xi\sim 1.7$, an absorber of velocity $-6000$~km~s$^{-1}$ (in the rest frame of the source) will show absorption at $6.42$~keV \textit{and} $7.0$--$8.0$~keV. 
 The warm absorber is required to have high column density ($N_{\rm H}\sim10^{23}$~cm$^{-2}$) in order to predict strong absorption in the Fe K edge region, or it may be that there is a super-solar abundance of iron.

(2) If the absorber is more ionized, for example at $\log\xi=2.6$, then an absorber velocity of $\sim 0$~km~s$^{-1}$ (with respect to the source) is needed to predict features at $7.0$--$8.0$~keV and $6.42$~keV. Here then, this may be just a high $\xi$ component to WA1 rather than a different absorber with a different velocity. Again we require a column density of $N_{\rm H}\sim10^{23}$~cm$^{-2}$.

(3) For ionizations $\log\xi$, between $1.7$ and $2.6$, the velocity shift will be somewhere between 
$-6000$~km~s$^{-1}$ and $0$~km~s$^{-1}$. Again, column densities need to be high, $N_{\rm H}\sim10^{23}$~cm$^{-2}$, to produce strong absorption features, or there is a super-solar abundance of Fe. High velocity dispersion could also help explain the broad trough. 

From situations (1)-(3), we present the possible parameters that do well at explaining the hard X-ray ($>5$~keV) spectrum in Table~\ref{abs2}. We are not able to deduce from the soft X-ray ($<2$~keV) spectrum which absorber is the best for the following reason.
The higher ionization models, ionization similar to that of (2), (which predict few soft X-ray lines) do not incorrectly predict strong lines in the soft X-ray but may incorrectly predict weak ones (the strength of predicted lines are less than noise in spectrum) where none is observed, while the lower ionization models, ionization similar to that of (1),  with redshift and ionization parameter similar to WA1 end up predicting many of the same lines as WA1, although not as well (partially due to the high column density required to predict strong absorption in the hard X-ray). It could be possible that there is only a single warm absorber component (WA1) with a super-solar abundance of Fe: if we increase Fe to about $10$ times the solar abundance WA1 shows strong absorption to explain the features in the vicinity of the Fe K edge.
However, a very high Fe abundance would imply lots of Fe L lines at lower energies, which are not seen. Additionally, the strength of the Fe K emission line  does not support a significantly high Fe abundance in the surrounding absorber.

(4) We also investigated the possibility that the absorption is due to absorption by the interstellar medium. In this case, we fixed the ISM absorber ({\sc ismabs}) redshift at $z=0$, and varied photoionization and column density to describe absorption at $6.42$~keV and $7.0$--$8.0$~keV.
However, again we find that we require a high column density  ($N_{\rm H}\sim10^{23}$~cm$^{-2}$, larger than what would be expected if we are looking through interstellar medium of the Galaxy), and so absorption around the source is preferred instead.

\begin{figure}[htbp]
\centering
\includegraphics[height=1.0\textwidth,angle=270.0]{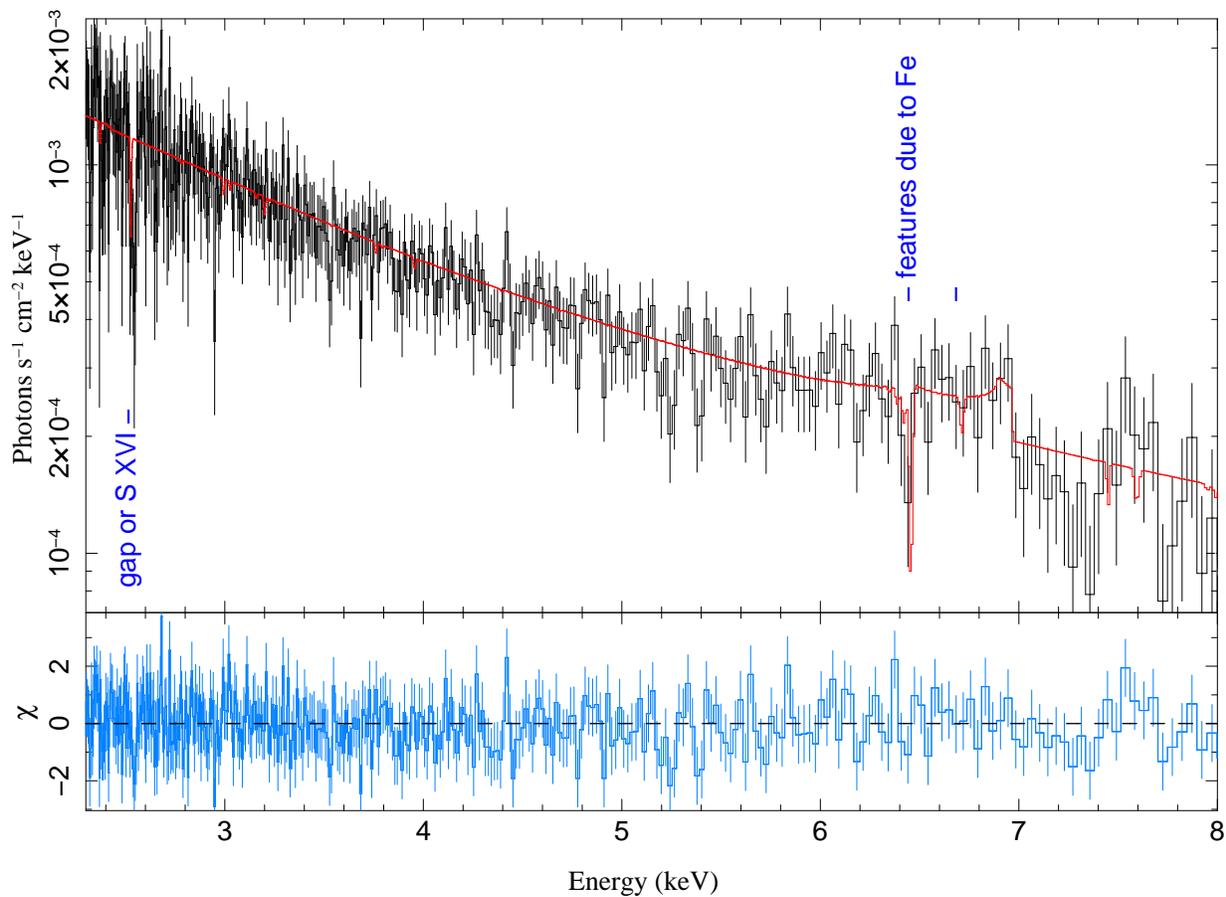}
\caption{A possible redshifted warm absorber? It may be possible that some of the strong absorption features in the hard X-ray spectrum are due to an ionized redshifted warm absorber, $\log\xi\sim 3$, $v\sim +5300$~km~s$^{-1}$. Such an absorber would also describe the strong absorption feature at $2.54$~keV, which may or may not be real due to a limited effective area (a possible gap in the chip) in the $2.5$--$2.7$~keV range in the $+1$ order spectra.}
\label{5300}
\end{figure}

(5) We investigated the case that the features are due to a redshifted warm absorber as well. 
We are unable to find reasonable parameters to predict strong absorption at $7.0$--$8.0$~keV. 
The $6.42$~keV absorption feature can be explained, accounted for with redshifted warm absorber models with $\log\xi\geq 2.9$ that are highly redshifted with respect to the source (relative velocity of $>5000$~km~s$^{-1}$) (see Fig.~\ref{5300}). 
Such a model could also explain the ambiguous absorption at $2.54$~keV mentioned in \S~\ref{NarrowAbsorptionLines}, which would mean that the $2.54$~keV is possibly real and is not due to a gap in the detection, however the model also incorrectly predicts strong Si lines. 
It is generally not easy to observe such highly redshifted absorbers, also disfavoring the redshifted model.



From all of the scenarios examined ((1)-(5)), the most likely explanation for the features in the vicinity of the Fe~K edge is a high ionization ($\log\xi\sim 2.3$ to $2.6$), high column density ($N_{\rm H}\sim10^{23}$~cm$^{-2}$) absorber (which has outflow velocity $\sim -3000$~km~s$^{-1}$ to $0$~km~s$^{-1}$ with respect to the source).
It is unlikely that the second warm absorber would have a lower ionization than WA1 but have a much higher column density. 
We need a high enough ionization parameter ($\log\xi > 2.3$) such that the Fe~K lines are still present but most of the lower Z elements (and Fe L shells) are virtually all fully stripped and therefore do not contribute to the absorption. The warm absorber with $\log\xi\sim 2.3$ also does the best job to reduce the Cash-statistic
(the change is $\sim 7$ for $3$ d.o.f., $P$-value $0.07$).

\section{Discussion and Conclusions}\label{DiscussionandConclusions}

The continuum of IRAS\,18325-5926 is best modeled by a power-law photon-index $\Gamma=2.01^{+0.07}_{-0.10}$ modified by a partial absorber of $N_{\rm H}=1.33^{+0.02}_{-0.03}\times10^{22}$~cm$^{-2}$ and $0.94^{+0.01}_{-0.01}$ covering fraction, and a line-of-sight Galactic column of $N_{\rm H}=7.4\times 10^{20} $~cm$^{-2}$. 
In addition, we use a $-360^{+41}_{-66}$~km~s$^{-1}$ warm absorber (WA1) with $\log\xi=2.01^{+0.07}_{-0.1}$ and $N_{\rm H}=1.6\times 10^{21}$~cm$^{-2}$, which describes the most prominent absorption features.
We also model the broad Fe K$\alpha$ emission as due to a disk at an inclination of ${25.8^\circ}^{+3.5}_{-7.6}$ 
with a simple diskline fit, or inclination $21.0^{+7.0}_{-21.0}$ with the blurred, ionized reflection model: {\sc reflionx}. 
A possible second warm absorber component is needed to describe some of the strong absorption 
due to  the complex profiles of iron K absorption in the vicinity of the Fe K edge. 
The model suggests that our line of sight grazes the edge of the obscuring torus of IRAS\,18325-5926 (if the source has one) or the source has a patchy global covering and we are able to see the regions affected by strong gravity near the SMBH as well, which would explain why IRAS\,18325-5926 has both Seyfert 1 and Seyfert 2 characteristics.

The iron K$\alpha$ emission line seen in the high resolution \textit{Chandra} spectrum of  IRAS\,18325-5926 is likely due to Fe XXV in a highly ionized accretion disk, in agreement with the conclusion of I04 about the origin of the emission feature.

We observe strong ionized ($\log\xi\sim 2$) blueshifted absorption features indicative of an X-ray outflow in IRAS\,18325-5926.
We modeled the \textit{Chandra} HETGS spectrum with a blueshifted $-360^{+41}_{-66}$~km~s$^{-1}$ warm absorber (WA1) with respect to the systemic velocity. 
The errors on the velocity, reported at $90\%$ confidence, seem to suggest that the warm absorber does have a relative velocity with respect to the central source, indicating that it may be an outflow. 
In the optical, a $-160^{+41}_{-66}$~km~s$^{-1}$ outflow is detected in the emission lines of [OIII] and the Balmer lines \citep{1995AJ....110..551I}. It is possible that the two outflows have a common origin.
There is the small chance that the warm absorber is not outflowing and rather the redshift of IRAS\,18325-5926 (considered to be $0.0198$) is actually slightly lower (closer to $0.0186$), although this is not likely. The redshift of IRAS\,18325-5926 is reported as $0.01982\pm0.00006$ based on narrow emission lines in the optical spectrum \citep{1995AJ....110..551I}, slightly higher
than the $z=0.0196$ value derived in \citet{1984AExpr...1...61C, 1985Natur.314..240D}.

We also detect absorption features (in the form of a broad trough) in the vicinity of the 
$>$7~keV iron K edge, likely due to the complex iron~K profiles noted by \citet{2004ApJS..155..675K}. 
A low number of counts and signal-to-noise in this region made it difficult quantify a good absorber model to explain this region.
However, as investigated and discussed in \S~\ref{VicinityoftheFeKEdge}, the most plausible explanation we find is a high ionization ($\log\xi\sim 2.3$ to $2.6$), high column density ($N_{\rm H}\sim10^{23}$~cm$^{-2}$), $v\sim -3000$ to $0$~km~s$^{-1}$ (with respect to source) absorber.

Given that absorber~$2$ cannot be well constrained, we consider the 
viewing geometry of IRAS\,18325-5926 in the context of the $-360$~km~s$^{-1}$ WA1 outflow. WA1 is not unusual compared to winds detected in other AGN (see \citealt{2005A&A...431..111B}).
The plasma properties of the absorber, described by $\log \xi$, can be
used to approximate the distance of the absorber from the source.
Following the calculations described in \citet{2002ApJ...567.1102L}, we take
\begin{equation}
\label{simple}
\xi=\frac{L_{\rm x}}{nR^2},
\end{equation}
where $R$ is the distance from the source of radiation causing photoionization, $L_{\rm x}$ is the X-ray luminosity, and $n=N_{\rm H} / \Delta  R$, with $N_{\rm H}$ being the hydrogen column of the absorber and $\Delta  R$ its thickness. In our model, we assume a value for the particle density of $n=10^4$~cm$^{-3}$. This is a plausible assumption but arbitrary (in so far as this is still an unknown) and the derived parameters are scaled by powers of $(n/10^4$~cm$^{-3})$. $R$ is found to be $1.4\,(n/10^4$~cm$^{-3})^{-1/2}$~pc.

We can then calculate the mass outflow rate due a detected wind assuming a spherical absorber with 
most of the mass of the absorber
with ionization parameter $\xi$ concentrated in a layer of thickness $\Delta R$ at distance $R$. The rate of outflow of material, given a wind speed of $v$ is
\begin{equation}
\label{outflow}
\dot M_{\rm wind}=4\pi R^2\rho v\left(\frac{\Omega}{4\pi}\right)=4\pi m_pv\left(\frac{L_{\rm x}}{\xi}\right)\left(\frac{\Omega}{4\pi}\right) C_\nu.
\end{equation}
The variable $\rho=\overline{n} m_{\rm p}$ is the density of the material in the absorber, where $m_{\rm p}$ is the mass of a proton and $\overline{n}$ is the average macroscopic ion number density. 
$\overline{n}$ is related to $n$, the microscopic electron density in the gas where
the physical absorption is taking place, by the volume filling factor, $C_\nu$, of the gas: $\overline{n}=n C_\nu$ (equation 15 of \citealt{2005A&A...431..111B}).
The volume filling factor of the gas cannot be directly measured and is difficult to estimate \citep{2007MNRAS.379.1359M}.
Therefore we use $C_\nu=1$ for simplicity.
We make a substitution to eliminate $R^2$ using Eq.~(\ref{simple}). $\Omega$ is the solid angle subtended by the outflow. Assuming a spherical outflow where $\Omega=4\pi$, Eq.~(\ref{outflow}) provides an upper limit on the mass outflow rate.
A summary of these calculations is shown in Table~\ref{outflowtbl}. As a check, $\Delta  R / R = 0.037$ for $(n=10^4$~cm$^{-3})$; the constraint that $\Delta  R / R \leq 1$ is met (see discussion in \citealt{2005A&A...431..111B} around equation 22). The dependence on $n$ is $\Delta  R / R = 0.037(n/10^4$~cm$^{-3})^{-1/2}$. 

The kinetic luminosity, $L_{\rm k}$ associated with a spherical mass outflow rate of $\dot M_{\rm wind}$ at velocity $v$ is
\begin{equation}
\label{kinetic}
L_{\rm k}=\frac{1}{2}\dot M_{\rm wind} v^2.
\end{equation}
The value of $L_{\rm k}$  can tell us how significant an outflow is in terms of energy.
For the $-360$~km~s$^{-1}$ wind, the kinetic luminosity is $8.6\times 10^{40}$~erg~s$^{-1}$. This power is only a small fraction $L_{\rm k}/L_{\rm x}\sim 0.005$ of the X-ray luminosity of the source.

We can also estimate the rate of accretion onto the black hole with
\begin{equation}
\label{accretion}
\dot M_{\rm accretion}=\frac{L_{\rm bol}}{\eta c^2}.
\end{equation}
The bolometric luminosity, $L_{\rm bol}$, can be approximated 
from the $2$--$10$~keV luminosity applying the bolometric correction of \citet{2004MNRAS.351..169M}.
For a Seyfert galaxy with a luminosity like IRAS\,18325-5926, the bolometric correction to the $2$--$10$~keV luminosity is about $10$, so we estimate that $\dot M_{\rm accretion}=2.2\times 10^{24}$~g~s$^{-1}=0.035 M_\odot$ yr$^{-1}$.
The rate of outflow due to the wind is about $2$ orders of magnitude greater than the accretion rate, if we assume the filling factor, $C_\nu$, is close to unity. Even if the filling factor is as small as $0.01$, the mass outflow rate is comparable to the accretion rate. Namely, one might conclude
that a significant amount of the mass appears to be leaving the IRAS\,18325-5926 galactic nuclei compared to the matter being captured by the accretion disk, although these two flows may result from different mechanisms and have different mass reservoirs since the distance of the outflow from the source, $R$, is found to be large ($1.35\,\,\,(n/10^4$~cm$^{-3})^{-1/2}$~pc). 

The Eddington luminosity of the source 
is $L_{\rm edd}=1.25\times10^{38} (M/M_\odot)
=1.25\times10^{45}$~erg~s$^{-1}$, 
where $M$ is the mass of the object in solar mass units, for which we use a value of $\sim 10^7 M_\odot$ \citep[][I04]{2005Ap&SS.300...67L}. Then, the ratio $L_{\rm x} / L_{\rm edd}$ is equal to $0.16$, meaning that IRAS\,18325-5926 is only at a small fraction of its Eddington luminosity. 
Since the warm absorber has a significantly higher opacity than a totally ionized gas,
a wind may be radiatively driven even if the source is only at a small fraction of its Eddington luminosity (see, for example, the steady-state, radiatively driven model by \citealt{1995MNRAS.273.1167R}).

We can calculate the escape velocity at the predicted distance of the warm absorber from the source to determine whether it is possible for the material to be returned to the host galaxy, assuming that most of the mass in this region is due to the black hole. The escape velocity $v_{\rm esc}$ at a distance $R$ is
\begin{equation}
\label{vesc}
v_{\rm esc}=\sqrt{\frac{2GM}{R}},
\end{equation}
where $G$ is the gravitational constant and $M$ is the mass of the black hole.
The calculated escape velocity at the distance of the warm absorber is $252~(n/10^4$~cm$^{-3})^{-1/4}$~km~s$^{-1}$. This value is comparable to and self-consistent with our measured $360$~km~s$^{-1}$ for the ionized outflow. The outflow appears to be able to escape into the surrounding environment, so the wind seems to be able to replenish the surrounding environment.

The large value for $\dot M_{\rm wind}/\dot M_{\rm accretion}\sim 10^2~C_\nu$ suggests that IRAS\,18325-5926 could have a significant impact on the large-scale surrounding environment of the black hole (unless $C_\nu$ is very small). AGN winds may serve as mechanisms that evolve the galaxy's structure as well as regulate the black hole, by putting ISM into the host galaxy, enriching the IGM and ISM with metals and heating the surrounding material \citep{2007ApJ...659.1022K}. If we assume that the outflow rate has been constant over the lifetime of the black hole which has accreted to a mass of $\sim 10^7 M_\odot$, then we estimate that $\sim 10^{9} C_\nu M_\odot$ of material has been returned to the host galaxy. Though, it is important to note that the mass reservoirs of the inflow and outflow may be different.
In terms of energy, however, the ratio $L_{\rm k}/L_{\rm bol}=4\times 10^{-4}$ of the kinetic luminosity of the wind to the bolometric luminosity of the source suggests the energy fed back by the wind is not significant compared to the bolometric luminosity.

The detected ionized outflow is likely not a short-term process. We can estimate the mass of the absorber as
\begin{equation}
\label{sphere}
M_{\rm abs}=\frac{4}{3}\pi \left((R+\Delta R)^3-R^3\right) \rho
\end{equation}
and obtain a characteristic timescale for the length of time the outflow will last 
\begin{equation}
\label{chart}
t=\frac{M_{\rm abs}}{\dot M_{\rm wind}}=\frac{(R+\Delta R)^3-R^3}{ 3R^2 v\left(\frac{\Omega}{4\pi}\right)}.
\end{equation}
The wind is expected to only last for approximately $140$~years (for $\Omega=4\pi$, $n=10^4$~cm$^{-3}$.
)  if there is no mechanism that feeds/replenishes the warm absorber column with matter. This is too short of a time scale to be reasonable as outflows are found in a fair number of AGN observed with high-resolution X-ray.  In a uniform analysis of high spectral resolution X-ray observations by \textit{Chandra} of $15$ Seyfert type galaxies, the study by \citet{2007MNRAS.379.1359M} found that $2/3$ of the AGN show signatures of blueshifted ionized absorbers. Outflows are widespread and thus likely have lifetimes comparable to the timescales of the accreting black hole.

The detected wind in IRAS\,18325-5926 likely originates in an obscuring torus structure, or the obscuring gas structure around the black hole:\citet{1995AJ....110..551I} suggests that the obscuring gas surrounding IRAS\,18325-5926 may be in the form of a global covering, rather than a toroidal distribution. \citet{2005A&A...431..111B} describes two different basic types of warm absorbers: torus winds seen in nearby Seyferts and accretion disk winds seen in quasars. 
We can estimate the inner radius of the obscuring torus around the black hole with
\begin{equation}
\label{torus}
r_{\rm inner}=L_{{\rm ion,}44}^{0.5},
\end{equation}
according to \citet{2001ApJ...561..684K}, where $L_{{\rm ion,}44}$ is the $1$--$1000$ Rydberg luminosity in units of $10^{44}$ erg~s$^{-1}$. We calculate that the inner radius of the torus is approximately at $4.3\times 10^{18}$~cm, which agrees well with the estimated distance of the low-velocity blueshift warm absorber ($4.1\times 10^{18}$~cm). Hence, the $-360$~km~s$^{-1}$ wind likely is a photoionized evaporation from the inner edge of the torus, or, if IRAS\,18325-5926 does not have a torus but rather a global covering gas as suggested in \citet{1995AJ....110..551I}, then the wind could possibly be launched from the outer edge of the accretion disk, or the surrounding global covering gas. 

Absorber~$2$ (with velocity $\sim -3000$~km~s$^{-1}$, $\log\xi=2.3$), if real, has a predicted distance of $0.7\,(n/10^4$~cm$^{-3})^{-1/2}$~pc, half the distance of WA1. Thus, likely this material would also be found around the torus or the outer edge of the accretion disk rather than the inner edge of the accretion disk. 

As an additional calculation, we can see how many gravitational radii the absorber WA1 is from the source. The gravitational radius of the black hole is
\begin{equation}
\label{genrel}
R_{\rm g}=\frac{GM}{c^2}{.}
\end{equation}
We find that $R_{\rm g}=1.5\times 10^{12}$~cm. Twice this value gives the Schwarzschild radius for a non-spinning black hole, while the value itself describes the event horizon location of a maximally spinning black hole. The $-360$~km~s$^{-1}$ wind is located at $\sim 2.7 \times 10^{6} R_{\rm g}$ from the event horizon, again suggesting that the wind likely originates in the torus or obscuring gas rather than in the inner edge of the accretion disk.

As mentioned in \S~\ref{Introduction}, X-ray winds have mostly been observed in Seyfert $1$ type galaxies. According to \citet{2005A&A...431..111B}, observed warm absorbers in nearby Seyfert $1$ type galaxies most likely originate in outflows from the dusty torus (\citet{2007MNRAS.379.1359M} findings support this as well). The Unified Model of AGN predicts that type $1$ and $2$ Seyfert galaxies differ only by the viewing angle of the obscuring torus around the nucleus \citep{1993ARA&A..31..473A,2008MNRAS.383.1501M}. Seyfert $2$ type galaxies, which display only narrow emission features, are thought to be viewed close to edge-on, whereas Seyfert $1$ type galaxies, having both broad and narrow emission features, are viewed close to face-on. The detection of an X-ray wind (likely originating from the distance of the torus) in the Seyfert $2$ object IRAS\,18325-5926 shows both types of Seyfert galaxies can have winds. It may be conceivable therefore that winds can have large subtended solid angles rather than being restricted to have a certain fixed orientation with respect to the torus and subtend only a narrow range of angles, or at the very least that winds can be observable for many different viewing orientation of the torus.

The full spectral shape of IRAS\,18325-5926 may be helpful in determining the origin of the wind.
The ratio of UV to X-ray flux, if too low, means that the material in the outer layer
of the accretion disk is too highly ionized to be accelerated to produce a wind \citep{2005A&A...431..111B,2003ApJ...585..406P}.

The X-ray spectrum of IRAS\,18325-5926 allows us to probe both the kinematics of the surrounding gas and also the broad emission feature thought to arise from an ionized accretion disk very near to the central source. It is possible that the surrounding gas of IRAS\,18325-5926 is patchy (in the case of a global covering model, with the wind possibly coming off the outer edge of the accretion disk), or that we are viewing the system at an angle just grazing the obscuring torus (if the source has one) thereby allowing us to see the nuclear region.

\acknowledgments

Acknowledgments: We thank Andrew Fabian for useful discussions and the referee for valuable comments. PM acknowledges Harvard University for the award of funding through the Harvard College Program for Research in Science and Engineering. JCL thanks the Harvard Faculty of Arts and Sciences and the Harvard College Observatory.

\newpage
\bibliography{mybib}{}
\bibliographystyle{plain}
\clearpage

\begin{deluxetable}{lcr}
\tabletypesize{\scriptsize}
\tablecaption{Best-fit parameters used to model the IRAS\,18325-5926 spectrum.}
\tablecolumns{3}
\tablewidth{0pt} 
\tablehead{\colhead{Component} & \colhead{Parameter} & \colhead{Value}}
\startdata
\multirow{3}{*}{Power-Law} & 	$\Gamma$ & $2.01^{+0.07}_{-0.10}$ \\
      & Norm (photon cm$^{-2}$ s$^{-1}$) & $0.010^{+0.001}_{-0.002}$ \\
      & $z$ & $0.0198$ \\ 
      \noalign{\smallskip} \tableline \noalign{\smallskip}
\multirow{2}{*}{Partial Cover}& $N_{\rm H}$ ($\times 10^{22}$ cm$^{-2}$) & $1.33^{+0.02}_{-0.03}$ \\
      & Cover fraction & $0.94^{+0.01}_{-0.01}$ \\
      \noalign{\smallskip} \tableline \noalign{\smallskip}
\multirow{1}{*}{Line-of-Sight Hydrogen Column} & $N_{\rm H}$ ($\times 10^{20}$ cm$^{-2}$) & $7.4$ \\
      \noalign{\smallskip} \tableline \noalign{\smallskip}
\multirow{6}{*}{Fe K$\alpha$ - Diskline} &	E (keV) & $6.73^{+0.06}_{-0.12}$ \\
      & Norm (photon cm$^{-2}$ s$^{-1}$) & $4.66^{+1.78}_{-2.65}\times10^{-5}$ \\ 
      & Power law dependence of emissivity & $-2.2^{+1.8}_{-0.9}$ \\
      & $R_{\rm in}$ ($GMc^{-2}$) & $6$ \\
      & $R_{\rm out}$ ($GMc^{-2}$) & $200$ \\
      & Inclination ($^{\circ}$) & $25.8^{+3.5}_{-7.6}$ \\
      \noalign{\smallskip} \tableline \noalign{\smallskip}
\multirow{4}{*}{Warm Absorber} 
      & $\log\xi$ & $2.01^{+0.06}_{-0.10}$ \\
      & Column ($\times 10^{21}$ cm$^{-2})$ & $1.55^{+0.75}_{-0.38}$ \\
      & $v_{\rm turb}$ (km s$^{-1}$) & $202^{+153}_{-61}$ \\
      & $v_{\rm wind}$ (km s$^{-1}$) (frame of IRAS 18325) & $-360^{+41}_{-66}$ \\
      \noalign{\smallskip} \tableline 
      \tableline
\sidehead{Model: tbabs(1)*zpcfabs(1)*warmabs(1)*(zpowerlw(1)+diskline(1))}  
\enddata
\label{par}
\end{deluxetable}

\begin{deluxetable}{lcrrrr}
\tabletypesize{\scriptsize}
\tablecaption{Models for the Fe K$\alpha$ emission line}
\tablecolumns{6}
\tablewidth{0pt} 
\tablehead{\colhead{Component} & \colhead{Parameter} & \colhead{Value} & \colhead{$\chi^2$} & \colhead{d.o.f.} & \colhead{$\chi_\nu^2$}}
\startdata
\multirow{6}{*}{Diskline} &	E (keV) & $6.73^{+0.06}_{-0.12}$ & $1194$ & $1130$ & $1.06$ \\
      & Norm (photon cm$^{-2}$ s$^{-1}$) & $4.66^{+1.78}_{-2.65}\times10^{-5}$  & & &  \\ 
      & $a^{\text{b}}$ & $2.2^{+1.8}_{-0.9}$  & & & \\
      & $R_{\rm in}$ ($GMc^{-2}$) & $6$ & & & \\
      & $R_{\rm out}$ ($GMc^{-2}$) & $200$ & & & \\
      & Inclination ($^{\circ}$) & $25.8^{+3.5}_{-7.6}$ & & & \\
      \noalign{\smallskip} \tableline \noalign{\smallskip}
\multirow{6}{*}{Laor} &	E (keV) & $6.73^{+0.12}_{-0.07}$ & $1197$ & $1129$ & $1.06$ \\
      & Norm (photon cm$^{-2}$ s$^{-1}$) & $1.33^{+0.33}_{-0.32}\times10^{-4}$ & & & \\ 
      & $a^{\text{b}}$ & $2.86^{+0.43}_{-0.32}$ & & & \\
      & $R_{\rm in}$ ($GMc^{-2}$) & $1.38^{+0.18}_{-0.14}$ & & & \\
      & $R_{\rm out}$ ($GMc^{-2}$) & $400$ & & & \\
      & Inclination ($^{\circ}$) & $26.0^{+4.8}_{-3.5}$ & & & \\
      \noalign{\smallskip} \tableline \noalign{\smallskip}
\multirow{6}{*}{Blurred, ionized reflection} &	$^1$reflionx: $\Gamma$ & $1.56^{+0.04}_{-0.04}$ & $1247$ & $1130$ & $1.10$ \\
      & reflionx: $\xi$ & $5174^{+50}_{-50}$ & & & \\
      & reflionx: Fe abund. ($\times$ Solar) & $1.0$ & & & \\
      & reflionx: norm (photon cm$^{-2}$ s$^{-1}$) & $2.56^{+0.05}_{-0.05}\times10^{-8}$ & & & \\
      & kdblur: $a^{\text{b}}$ & $1.97^{+0.12}_{-0.14}$ & & & \\
      & kdblur: $R_{\rm in}$ ($GMc^{-2}$) & $1.24^{+2.06}_{-0.10}$ & & & \\
      & kdblur: $R_{\rm out}$ ($GMc^{-2}$) & $100$ & & & \\
      & kdblur: inclination ($^{\circ}$) & $21.0^{+7.0}_{-21.0}$ & & & \\
      & powerlw: norm (photon cm$^{-2}$ s$^{-1}$) & $1.77^{+0.08}_{-0.07}\times10^{-3}$ & & & \\
      & powerlw: $\Gamma$ & tied to $^1$ & & & \\
      & phabs: $N_{\rm H}$ ($\times10^{21} $~cm$^{-2}$) & $8.0^{+0.1}_{-0.2}$ & & & \\                 
      \noalign{\smallskip} \tableline  
\enddata
\tiny
\tablenotetext{b}{Power law dependence of emissivity $a$ (scales as $R^{-a}$) }
\label{par2}
\end{deluxetable}

\begin{deluxetable}{llllrrrcc}
\tabletypesize{\tiny}
\tablecaption{Observed and Predicted Strong Absorption Lines due to Warm Absorber $1$ based on {\sc XSTAR}.}
\tablecolumns{9}
\tablewidth{0pt} 
\tablehead{\colhead{ID\#} & \colhead{Ion} & \colhead{Transition} & \colhead{$f_{ij}^a$} & \colhead{$\lambda_{lab}^b$} & \colhead{$\lambda_{obs}^c$} & \colhead{$\tau^d$} & \colhead{$W_{\lambda}^e$} & \colhead{H/MEG}
}
\startdata
$1$ & O {\sc viii} & Ly $\beta$:  $1s(^2S)\rightarrow 3p(^2P^o)$ & 0.079 & 16.006 & 16.304 & 362.00 & 30.00  & m \\
$2$ & O {\sc viii} & Ly $\gamma$:  $1s(^2S)\rightarrow 4p(^2P^o)$ & 0.029 & 15.188 & 15.471 & 133.00 & 17.00  & m \\
$3$ & O {\sc viii} & Ly $\delta$:  $1s(^2S)\rightarrow 5p(^2P^o)$ & 0.014 & 14.832 & 15.108 & 61.00 & 9.80  & m \\
$4$ & O {\sc viii} & Ly $\theta$:  $1s(^2S)\rightarrow 6p(^2P^o)$ & 0.008 & 14.645 & 14.917 & 33.60 & 5.90  & m \\
$5$ & Ne {\sc ix} & He $\alpha$:  $1s^2(^1S_{0})\rightarrow 1s2p(^1P_{1}^o)$ & 0.721 & 13.447 & 13.697 & 272.00 & 22.00  & m \\
$6$ & Ne {\sc ix} & He $\beta$:  $1s^2(^1S_{0})\rightarrow 1s3p(^1P_{1}^o)$ & 0.148 & 11.547 & 11.762 & 48.10 & 6.20  & m \\
$7$ & Ne {\sc x} & Ly $\alpha$:  $1s(^2S_{1/2})\rightarrow 2p(^2P^o)$ & 0.415 & 12.134 & 12.360 & 445.00 & 25.00  & m \\
$8$ & Ne {\sc x} & Ly $\beta$:  $1s(^2S_{1/2})\rightarrow 3p(^2P^o)$ & 0.079 & 10.240 & 10.431 & 85.60 & 8.50  & h \\
$9$ & Mg {\sc xi} & He $\alpha$:  $1s^2(^1S_{0})\rightarrow 1s2p(^1P_{1}^o)$ & 0.738 & 9.169 & 9.340 & 192.00 & 13.00  & h \\
$10$ & Mg {\sc xi} & He $\beta$:  $1s^2(^1S_{0})\rightarrow 1s3p(^1P_{1}^o)$ & 0.151 & 7.851 & 7.997 & 33.90 & 3.10  & h \\
$11$ & Mg {\sc xii} & Ly $\alpha$:  $1s(^2S_{1/2})\rightarrow 2p(^2P^o)$ & 0.414 & 8.421 & 8.578 & 108.00 & 8.30  & h \\
$12$ & Si {\sc xiii} & He $\alpha$:  $1s^2(^1S_{0})\rightarrow 1s2p(^1P_{1}^o)$ & 0.748 & 6.648 & 6.772 & 186.00 & 9.40  & h \\
$13$ & Si {\sc xiv} & Ly $\alpha$:  $1s(^2S_{1/2})\rightarrow 2p(^2P^o)$ & 0.414 & 6.182 & 6.297 & 31.70 & 2.50  & h \\
$14$ & S {\sc xv} & He $\alpha$:  $1s^2(^1S_{0})\rightarrow 1s2p(^1P_{1}^o)$ & 0.761 & 5.039 & 5.133 & 52.80 & 3.50  & h \\
$15$ & Fe {\sc xvii} & $2s^22p^6(^1S_{0})\rightarrow 2s^22p^53s(^1P_{1})$ & 0.122 & 17.050 & 17.367 & 20.50 & 4.10  & m \\
$16$ & Fe {\sc xvii} & $2s^22p^6(^1S_{0})\rightarrow 2s^22p^53d(^3D_{1})$ & 0.596 & 15.262 & 15.546 & 89.60 & 13.00  & m \\
$17$ & Fe {\sc xvii} & $2s^22p^6(^1S_{0})\rightarrow 2s^22p^53d(^1P_{1})$ & 2.517 & 15.015 & 15.294 & 330.00 & 28.00  & m \\
$18$ & Fe {\sc xvii} & $2s^22p^6(^1S_{0})\rightarrow 2s2p^63p(^1P_{1})$ & 0.283 & 13.823 & 14.080 & 38.00 & 5.90  & m \\
$19$ & Fe {\sc xvii} & $2s^22p^6(^1S_{0})\rightarrow 2s^22p^54d(^3D_{1})$ & 0.374 & 12.264 & 12.492 & 44.50 & 6.00  & m \\
$20$ & Fe {\sc xvii} & $2s^22p^6(^1S_{0})\rightarrow 2s^22p^54d(^1P_{1})$ & 0.434 & 12.123 & 12.349 & 48.30 & 6.70  & m \\
$21$ & Fe {\sc xviii} & $2s^22p^5(^2P_{3/2})\rightarrow 2s^22p^43d(^4P_{3/2})$ & 0.104 & 14.549 & 14.820 & 21.30 & 4.10  & m \\
$22$ & Fe {\sc xviii} & $2s^22p^5(^2P_{3/2})\rightarrow 2s^22p^43d(^2F_{5/2})$ & 0.203 & 14.537 & 14.807 & 46.00 & 7.40  & m \\
$23$ & Fe {\sc xviii} & $2s^22p^5(^2P_{3/2})\rightarrow 2s^22p^43d(^2D_{5/2})$ & 0.311 & 14.376 & 14.643 & 66.70 & 10.00  & m \\
$24$ & Fe {\sc xviii} & $2s^22p^5(^2P_{3/2})\rightarrow 2s^22p^43d(^2S_{1/2})$ & 0.230 & 14.258 & 14.523 & 49.50 & 7.90  & m \\
$25$ & Fe {\sc xviii} & $2s^22p^5(^2P_{3/2})\rightarrow 2s^22p^43d(^2P_{3/2})$ & 0.590 & 14.208 & 14.472 & 127.00 & 16.00  & m \\
$26$ & Fe {\sc xviii} & $2s^22p^5(^2P_{3/2})\rightarrow 2s^22p^43d(^2D_{5/2})$ & 0.937 & 14.206 & 14.470 & 208.00 & 21.00  & m \\
$27$ & Fe {\sc xviii} & $2s^22p^5(^2P_{3/2})\rightarrow 2s^22p^43d(^2D_{3/2})$ & 0.127 & 14.155 & 14.418 & 27.10 & 4.60  & m \\
$28$ & Fe {\sc xix} & $2s^22p^4(^3P_{2})\rightarrow 2s^22p^33d(^3D_{3})$ & 0.220 & 13.799 & 14.056 & 75.50 & 10.00  & m \\
$29$ & Fe {\sc xix} & $2s^22p^4(^3P_{2})\rightarrow 2s^22p^33d(^3F_{3})$ & 0.087 & 13.648 & 13.902 & 29.60 & 4.70  & m \\
$30$ & Fe {\sc xix} & $2s^22p^4(^3P_{2})\rightarrow 2s^22p^33d(^3P_{2})$ & 0.120 & 13.555 & 13.807 & 40.40 & 6.10  & m \\
$31$ & Fe {\sc xix} & $2s^22p^4(^3P_{2})\rightarrow 2s^22p^33d(^3D_{3})$ & 0.748 & 13.525 & 13.777 & 250.00 & 21.00  & m \\
$32$ & Fe {\sc xix} & $2s^22p^4(^3P_{2})\rightarrow 2s^22p^33d(^3D_{2})$ & 0.374 & 13.506 & 13.757 & 123.00 & 15.00  & m \\
$33$ & Fe {\sc xix} & $2s^22p^4(^3P_{2})\rightarrow 2s^22p^33d(^3S_{1})$ & 0.252 & 13.456 & 13.706 & 84.10 & 11.00  & m \\
$34$ & Fe {\sc xix} & $2s^22p^4(^3P_{2})\rightarrow 2s^22p^33d(^1F_{3})$ & 0.204 & 13.430 & 13.680 & 61.50 & 9.30  & m \\
$35$ & Fe {\sc xix} & $2s^22p^4(^3P_{2})\rightarrow 2s^22p^34d(^3D_{3})$ & 0.137 & 10.816 & 11.017 & 36.30 & 4.50  & m \\
$36$ & Fe {\sc xx} & $2s^22p^3(^4S_{3/2})\rightarrow 2s^22p^23d(^4P_{5/2})$ & 0.501 & 12.845 & 13.084 & 42.20 & 6.60  & m \\
$37$ & Fe {\sc xx} & $2s^22p^3(^4S_{3/2})\rightarrow 2s^22p^23d(^4P_{3/2})$ & 0.505 & 12.827 & 13.066 & 46.70 & 6.60  & m \\
\enddata
\tiny
\tablenotetext{a}{~Oscillator Strength.}
\tablenotetext{b}{~Laboratory Wavelength.}
\tablenotetext{c}{~Observed Wavelength.}
\tablenotetext{d}{~Optical Depth.}
\tablenotetext{e}{~Equivalent Width.}
\label{lines}
\end{deluxetable}

\begin{deluxetable}{lcr}
\tabletypesize{\scriptsize}
\tablecaption{Possible second warm absorber parameters that describe the hard X-ray ($>2$)~keV features}
\tablecolumns{2}
\tablewidth{0pt} 
\tablehead{\colhead{Ionization ($\log\xi$)} & \colhead{Redshift $z$}} 
\startdata
$1.7$ & $0$ \\
$2.0$ & $0.006$ \\
$2.3$ & $0.01$ \\
$2.6$ & $0.0198$ 
\enddata
\label{abs2}
\end{deluxetable}

\begin{deluxetable}{ccccc}
\tabletypesize{\normalsize}
\tablecaption{Warm absorber properties.}
\tablecolumns{5}
\tablewidth{0pt} 
\tablehead{\colhead{$R$} & \colhead{$\Delta R$} & \colhead{$\Delta R/R$}  & \colhead{$\dot M_{\rm wind}$ } & \colhead{$L_{\rm k}$ } 
\\
\colhead{($(n/10^4$~cm$^{-3})^{-1/2}$ cm)} & \colhead{$(n/10^4$~cm$^{-3})^{-1}$ cm)} & \colhead{($(n/10^4$~cm$^{-3})^{-1/2}$)} & \colhead{($M_\odot$ yr$^{-1}$)} & \colhead{(erg~s$^{-1}$)} 
}
\startdata
$4.18\times10^{18}$ & $1.55\times10^{17}$ & $0.037$ & $2.1$ & $8.6\times 10^{40}$\\
\enddata
 \label{outflowtbl}
\end{deluxetable}

\end{document}